Loki, Io: New groundbased observations and a model describing the change from periodic overturn

J. A. Rathbun – Department of Physics, University of Redlands, 1200 E. Colton Ave., Redlands, CA 92374 (julie_rathbun@redlands.edu)

J. R. Spencer – Southwest Research Institute, 1050 Walnut St., Suite 400, Boulder, CO 80302 (spencer@boulder.swri.edu)

ABSTRACT

Loki Patera is the most powerful volcano in the solar system.  We have obtained measurements of Loki's 3.5 micron brightness from NASA's Infrared Telescope Facility (IRTF) and have witnessed a change from the periodic behavior previously noted.  While Loki brightened by a factor of several every 540 days prior to 2001, from 2001 through 2004 Loki remained at a constant, medium brightness.  We have constructed a quantitative model of Loki as a basaltic lava lake whose solidified crust overturns when it becomes buoyantly unstable.  By altering the speed at which the overturn propagates across the patera, we can match our groundbased brightness data.  In addition, we can match other data taken at other times and wavelengths.  By slowing the propagation speed dramatically, we can match the observations from 2001-2004.  This slowing may be due to a small change in volatile content in the magma.

1. Introduction

Loki Patera is a horseshoe-shaped dark feature 200 km across located on the Jupiter-facing hemisphere of the volcanically active moon Io.  At more than 200 km across, it is the largest volcanic feature on Io and is the site of large amounts of thermal emission, at times accounting for nearly 15% of Io's total heat flow (Spencer et al., 2000).  While its appearance at visible wavelengths has changed little since it was observed by the Voyager spacecraft (McEwen et al., 1998),  its near infrared brightness can vary by an order of magnitude over a timescale of several months.

1.1. Telescopic Observations

Loki's brightness in the infrared is large enough to be measured with ground-based telescopes using occultation techniques (Spencer et al., 1990).  In this technique, Io is viewed while in eclipse, thereby eliminating reflected sunlight from the brightness measurement, and undergoing an occultation by Jupiter.  As the limb of Jupiter covers (or uncovers) Io, active hotspots disappear from Io's total brightness.  Io's total brightness is measured photometrically as a function of time and each hotspot appears as a step in the resulting lightcurve.  Individual hotspot brightnesses can be determined from the height of the step.  Loki's brightness has been measured in this manner for more than a decade (Rathbun et al., 2002).  Its variability, indicative of fluctuations in the high-temperature component, is apparent.  Rathbun et al. (2002) tested the variability and found that the brightness from 1988 through 2001 varied periodically, with a period of 540 days.

We have continued to measure Loki's brightness from NASA's Infrared Telescope Facility (IRTF) using the occultation technique (Fig. 1).  Although the temporal sampling rate has decreased, a change in Loki's behavior clearly occurred in mid-2001.  Data taken before this time show two distinct populations of brightness: one low and one high. Data obtained between late 2001 and early 2003 show a single brightness population at a level between the previous "high" and "low" levels. In late 2003 the brightness fades to a low level and increases to a high level, indicating that another brightening event has begun. This new event appears to be out of phase with the earlier brightening events, but the temporal sampling has been inadequate recently to fully characterize Loki's behavior.

1.2. Spacecraft Data

From 1999 through 2001 the Galileo spacecraft performed a series of close flybys of Io and obtained high resolution images from multiple remote sensing instruments. Spencer et al. (2000) examined high spatial resolution temperature measurements of Loki taken by the Galileo PhotoPolarimeter-Radiometer (PPR) in October 1999 and February 2000. The PPR instrument is sensitive from visible wavelengths up to 100 micron infrared wavelengths. Observations of Loki were taken at night to avoid reflected sunlight so the brightness measurements are dominated by low-temperature ($\leq 300$ K) emission. They found that the hottest part of Loki moved from the southwest corner in 1999 to the eastern edge in 2000. In the October 1999 data the temperature gradually increased from the southwest corner until it abruptly drops. Rathbun et al. (2002) interpreted this as evidence of a front of hot material moving to the east. They further noted that Voyager images taken in March and July of 1979 also show evidence of a front of dark material moving to the east. In both cases, they interpret the moving material to be young. The hot material is obviously young, while the presence of an active plume near Loki during the Voyager flyby depositing bright material suggests that the dark material is young.

Rathbun et al. (2002) favor the interpretation of Loki as an overturning lava lake. In order to match these data, they suggest that a front of crustal foundering moves across Loki. As the crust cools and thickens by freezing of lava on its base, it eventually becomes negatively buoyant. One piece sinks, causing the piece next to it to sink and thus the resurfacing progresses in a coherent manner. So, the Galileo PPR data, which capture the low-temperature emission, show the warmer part of Loki that has already overturned and the cooler temperatures to the east where older material not yet overturned. Davies (2003) analyzed a high resolution infrared image from the Near Infrared Mapping Spectrometer (NIMS). Although the front of the foundering was not seen in the image, the surface ages inferred from the observation varied nearly linearly across Loki decreasing to the northeast, consistent with the overturn of a lava lake in the manner proposed by Rathbun et al. (2002).

2. Model

Here, we quantify the model of Rathbun et al. (2002) and determine whether, in addition to matching the PPR data, it can quantitatively match the high-temperature emission that dominates the integrated 3.5 micron flux that is measured from the ground and if it can yield insight into the behavior changes observed in those data.

Loki Patera has an area of approximately $2.1 \times 10^4$ km$^2$ and a width of 55 km. For simplicity, we model Loki as a rectangular lava lake 390 km long and 55 km wide (as if the horseshoe of Loki were straightened; figure 2). The depth of the lava lake is not set, but we assume it is deep enough that the bottom does not affect our calculations. We model the age of the surface of the lake as a function of position and time. Only the long dimension of position matters and it is broken into a large number, n, of rafts. Each raft acts as a single unit and will overturn as one piece. The model begins as a new foundering front begins to propagate from the west end of the Patera. The age of the surface, created in the previous wave of foundering, varies linearly along the length of the Patera, with the oldest surface in the west. Based on the early groundbased observations (Rathbun et al., 2002), we set the oldest crust at 540 days and the youngest at 150 days. In time, the rafts in the west begin to founder and the foundering front moves to the east. No raft founders until its western neighbor has already foundered. The major variables in this model are the speed with which the foundering progresses and the total number (and therefore size in the East/West direction) of rafts. Based on these two variables, we calculate the number of rafts which will overturn in one day. When rafts overturn, their ages are set to less than one day old, with the exact age varying linearly from one day in the west to zero days in the east. All rafts that do not overturn are aged by one day. While the model using a time increment of one day, the results are the same as for a continuous process of overturn.

Once the age of the surface is calculated as a function of distance and time, we then use the lava cooling model of Davies et al. (2005) to calculate the temperature of the surface. This finite-element model assumes a basaltic composition of the lava with a liquidus temperature of 1200-1600 K. Finally, knowing the area and surface temperature of each raft and assuming blackbody emission, we determine the total brightness of Loki at 3.5 microns, simulating what would be seen from the ground (most groundbased data record only Loki's integrated thermal emission).

For raft sizes larger than one hundred meters across (5.5 km$^2$ in area), the calculated brightness decreases as the raft size increases. As the raft size increases, the number of rafts overturning in a day decreases. Since the age of the youngest raft is one divided by the number of rafts that overturn and it is this age that determines the hottest temperature, this lowers the maximum temperature, and thus brightness, reached. At sizes smaller than about 100 m across, the decrease in size of the raft balances the increase in temperature of the youngest raft and the model is independent of raft size and is indistinguishable from a continuous process. For simplicity, we begin with the approximately continuous case. This is a reasonable assumption since no rafts are resolved in the Galileo data. Since the calculation time for the model increases dramatically with decreasing raft size, we use a raft size of 2 m (i.e., a raft area of 0.1 km$^2$) in all subsequent calculations.

Altering the speed of the foundering front also affects the number of rafts that founder in a day. The faster it moves, the more surface area younger than a given age is produced, thus increasing the total brightness. We find that the brightness is proportional to velocity with the constant of proportionality approximately 32 when the brightness is measured in GW/micron/str and the propagation speed in km/day.

3. Model results

The groundbased data (figure 1) show very little fluctuation of brightness over short timescales (minutes to hours), so we infer that the foundering is a relatively steady process. We interpret changes in brightness at longer timescales (days to months) to indicate variations in the overturn propagation speed. Approximately 9 brightening events can be seen in the data, but only five have well defined starting and ending times. The average duration of these 5 brightenings is 225 days, requiring a propagation speed of 1.7 km/day. Given this speed, the model predicts an average active brightness of 55 GW/µm/str, very closed to the average active brightness of 60 GW/µm/str observed during these 5 events (figure 3b).

3.1 Sulfur

Voyager's Infrared Interferometer Spectrometer (IRIS) observed Loki's thermal emission at 5 - 50 microns: emission at these wavelengths could be fit with a maximum temperature of only 450 K (Pearl and Sinton 1982). This relatively low temperature led to the suggestion that Loki might be a sulfur lake (Lunine and Stevenson 1985), though the IRIS data have also be fitted with silicate models (Carr 1986, Howell 1997). For completeness, we checked whether our groundbased 3.5 micron data could be reconciled with a sulfur lake model.

While the Davies et al. (2005) cooling model for basalt is numerical, Howell (1997) constructs an analytic formulation in terms of the chemical properties of the magma. He finds that $T(t)=a_T t^{-1/8}$ where $a_T = \left( \frac{1}{\sqrt{\pi}} \left[ \frac{\sqrt{K\rho C}}{\mathrm{Erf}(\Lambda)} \right] \frac{\Delta T}{\sigma} \right)^{1/4}$, K is the thermal conductivity, ρ is the density, C is the heat capacity, σ is the Stefan-Boltzman constant, and Λ is the solution to the transcendental equation

$$\frac{L}{C\Delta T} = \frac{e^{-\Lambda^2}}{\Lambda\sqrt{\pi}\,\mathrm{Erf}(\Lambda)}$$

where L is the heat of fusion. Since sulfur is a single element, it's material properties are better constrained than those of basalt. We use values of K = 0.205 W/mK, ρ = 1.819 g/cm³, C = 22.75 J/(mol K), ΔT = 100 K, L = 1.727 kJ/mol, and an atomic mass of 32.065 g/mol to find that $a_T$ = 225.7 K. The maximum temperatures reached during our model are well below the boiling point of sulfur. With this modification to our model, we again attempted to match the 3.5 micron brightness of Loki. Even with ridiculously large speeds (10 km/day) the maximum brightnesses modeled were far below the brightnesses observed. Further, at those speeds, the overturn event lasts a significantly shorter period of time than observed, 39 days versus the more typical 230 to 390 days. Figure 3 shows the combinations of duration and speed that apply to Sulfur and those for Basalt. From this, we conclude that Loki patera is not an overturning lake of liquid sulfur.

3.2. Modeling individual brightenings, 1997-2000

The best temporal resolution data for Loki were taken during the period of 1997 through 2000, during the Galileo era. They showed that Loki's brightness variations are more complex than a simple on/off model. We therefore tested whether we could match the brightness variations observed by varying with time the propagation speed in the basaltic model. We used the simplest velocity profile possible, i.e. constant or changing linearly with time, and our simple model was able to match the observed brightnesses remarkably well. (figure 3a). The mean velocity for this modeled period is 1.53 km/day. Since the material overturning is 55 km wide, the rate at which new material is exposed at the surface is 970 m$^2$/s, comparable to the rate of 1160 m$^2$/s found by Howell et al. (2001). Howell was specifically looking at October and November of 1999, where our velocities are 1.8 km/day, equivalent to an exposure rate of 1150 m$^2$/s.

Using speckle techniques, MacIntosh et al. (2003) observed Io and were able to measure Loki's 2.2 micron brightness on July 12$^{th}$ and 28$^{th}$ and August 4$^{th}$, 1998. We ran the same model from the 1997 thru 2001 period, but calculating the 2.2 micron brightness instead. The model again matches the observed brightnesses to within 10%. The nearest 3.5 micron observations were taken June 19$^{th}$ and August 20$^{th}$, so a direct comparison of the model with multiple wavelengt6hs simultaneously is not possible. MacIntosh et al. (2003) were also able to determine the position of the Loki hot spot on July 12$^{th}$ and August 4$^{th}$, 1998. Using the same velocity profile from the model, we calculated the location of the overturning front on those dates to be 64 and 86 km from the western edge, respectively. To compare these locations with MacIntosh's observations we need to take our linear model and transform it back into a horseshoe shape. Figure 3c shows a white horseshoe representing the long dimension in our model with asterisks at the positions we calculated for the front on the two days. The background is an image from MacIntosh et al. (2003) showing a Voyager image of Loki and two error boxes (in black) for the position of the hottest area observed on those dates. Our modeled positions fall within these error boxes.

3.3 Modeling data from 2001

In 2001, Loki's 3.5 micron behavior appeared to change. No longer did the brightness alternate between "bright" and "dark". Instead, it remained at a reasonably constant 30-45 GW/micron/str for at least 500 and up to 900 days. A brightening event was observed to begin between September 15$^{th}$ and October 10$^{th}$, 1999. The brightening was still occuring on March 7$^{th}$, 2000. No data was taken between then and December 24$^{th}$, 2000. In that time, we infer that brightening event ended. Between December 24$^{th}$, 2000 and April 14$^{th}$, 2002 fifteen observations were made with an average brightness of 37 GW/micron/str and a smaller than normal deviation from this average. Loki's brightness then remained near the "dark" level until a new "bright" event began in late 2003, out of phase with the previous events.

Simply reducing the velocity of the overturn propagation allows our model to match the brightnesses observed in 2001-2002. A speed of 0.9 km/day yields an event with a maximum brightness of 29 GW/micron/str and a duration of 450 days, a reasonable fit to observations. This velocity is similar to that calculated by Davies (2003). He fit temperatures to high spatial resolution NIMS spectra of Loki taken October 10$^{th}$, 2001. Those data did not show the front of hot material, likely because the observation covered only the southern part of the patera, but from the temperature variation across the Patera, he calculated a resurfacing rate of 1 km/day.

The NIMS observation is consistent with our model (Davies, 2003) if the overturn front has already passed the area observed. In our model, the hot overturning material dominates the 3.5 micron flux. The total 3.5 micron brightness measured by NIMS was approximately 2 GW/micron/str. Since the observation covered roughly half the patera, we calculate an average brightness of the entire patera minus the overturn front to be approximately 4 GW/micron/str. Depending on the time elasped since the overturn front passed and the propagation speed, the model predicts a brightness of 0.5-5 GW/micron/str. When the front is present, the total modeled brightness is several tens of GW/micron/str, again depending on overturn propagation speed.

Observations of Io using adaptive optics can isolate the brightness from individual volcanoes. Marchis et al. (2005) measured Loki's brightness on December $18^{th}$, $20^{th}$, and $28^{th}$, 2001 at 2.2, 3.8, and 4.3 microns. These data show much more variability at 3.8 and 4.3 microns than is seen in the occultation data. While we were not able to model these short-term variations, by using a velocity of 0.5 km/day we are able to reasonably match the average brightnesses measured (figure 4a). This is slightly lower than the average value of 0.9 km/day used to match the occultation measurements, but the 1997-2000 data show that variations in velocity by a factor of 2 are not unusual.

When velocities are as low as 0.5 km/day, it takes 780 days to overturn the entire patera. We had previously assumed that a new brightening even began after 540 days when the solidified crust was again of sufficient density to founder. In the past, this new overturn always began after the previous one was finished. Furthermore, simply by changing the velocity to 0.5 km/day, the easternmost rafts are already older than 540 days when they sink, so it is not clear that a when a new overturn will begin. However, we model the case of simultaneous overturn fronts for completeness. We assume both fronts have a velocity of 0.5 km/day, both fronts begin at the western edge, and that the second front begins 540 days after the first (figure 4b). The maximum brightness occurs when both fronts are present and is 33 GW/micron/str, twice the 17 GW/micron/str when only one front is active. So, another way to match the observations from 2001 through 2002 is for there to be simultaneous fronts for part of the time, perhaps at different speeds. Without more data, we cannot unambiguously determine which mechanism accounts for the change in Loki's behavior, but a change is velocity is required in either case.

4. Discussion

We have shown that a simple quantitative model of an overturning basaltic lava lake is consistent with observations by a variety of observers at a variety of wavelengths using a variety of instruments and techniques including groundbased occultation measurements at 3.5 microns, groundbased speckle measurements at 2.2 microns plus position, groundbased adaptive optics measurements at 2.2, 3.8, and 4.3 microns, and Galileo PPR and NIMS data at mid-infrared and far-infrared wavelengths. All the data can be matched by simple changes in the velocity of the overturn propagation. But, why does the velocity change?

A change in velocity indicates a change in the age of the surface when it sinks. In the original Rathbun et al. (2002) model the rafts overturn when their density increases to the point where they are denser than the underlying liquid in the lake. In order for the solid crust to not sink immediately, there must be some porosity in the crust that lowers its density to lower than the liquid. Peck et al. (1966) measured the porosity of a solidified Hawaiian lava lake as a function of depth. We use a simple Stefan model to calculate the thickness of the solidified crust as a function of time and incorporate the Peck et al. (1966) measurements to calculate the density of the solidified crust as a function of time (figure 5). The density remains remarkably constant between approximately 400 and 800 days, with only a 1% variation over that time. This is due to the fact that the thickness changes little at this time and also the porosity changes little with depth, resulting in little variation of density at that time. This shows that small differences in density of the magma will yield large differences in the age of the raft when it founders. Further, small differences in initial density (especially porosity) of the crust can also yield large differences in age. Finally, other factors, such as the behavior of neighboring slabs, can also have an effect on when a particular raft sinks. We believe the most likely explanation for the variations in age at which the raft sinks is small changes in magma volatile content, which affects both the magma and crust density. These small changes can produce large variations in the sinking age and thus the propagation speed of the sinking front.

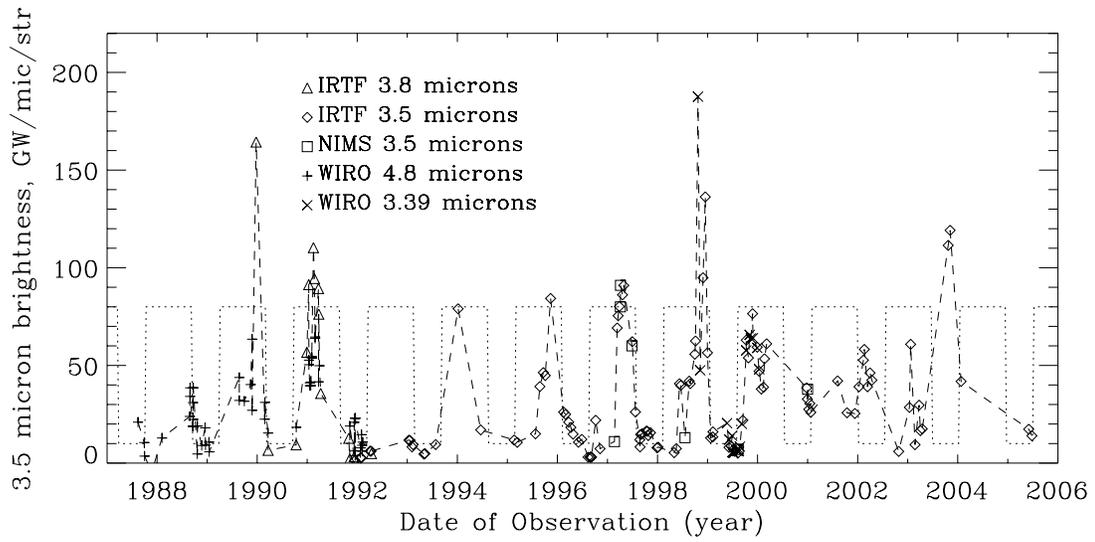

**Figure 1:** 3.5 μm brightness of Loki as measured primarily from Jupiter occultations. Some of the data was taken at other wavelengths (3.8, 4,8, and 3.39 μm), see Rathbun et al. (2002) for details. The dotted square wave has a period of 540 days to show the early periodicity.

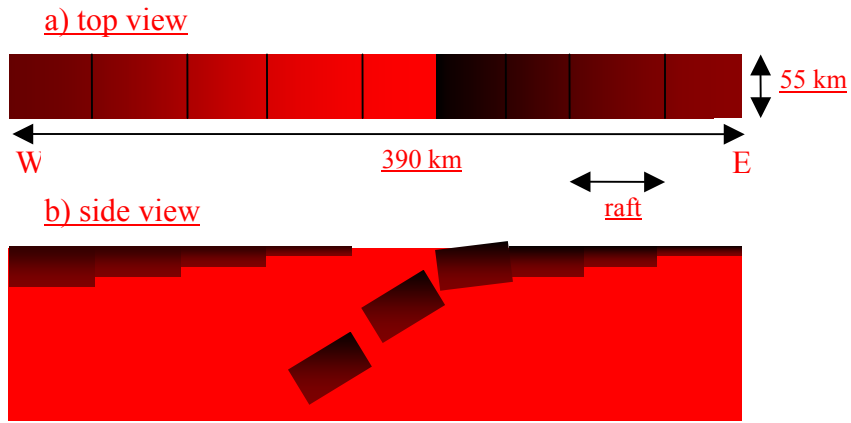

**Figure 2:** A graphical depiction of our model. The top view shows how temperature varies along the 390 km length of the lava length. The side view shows how the crust thickens with time until eventually sinking below the liquid magma and a new crust begins to form.

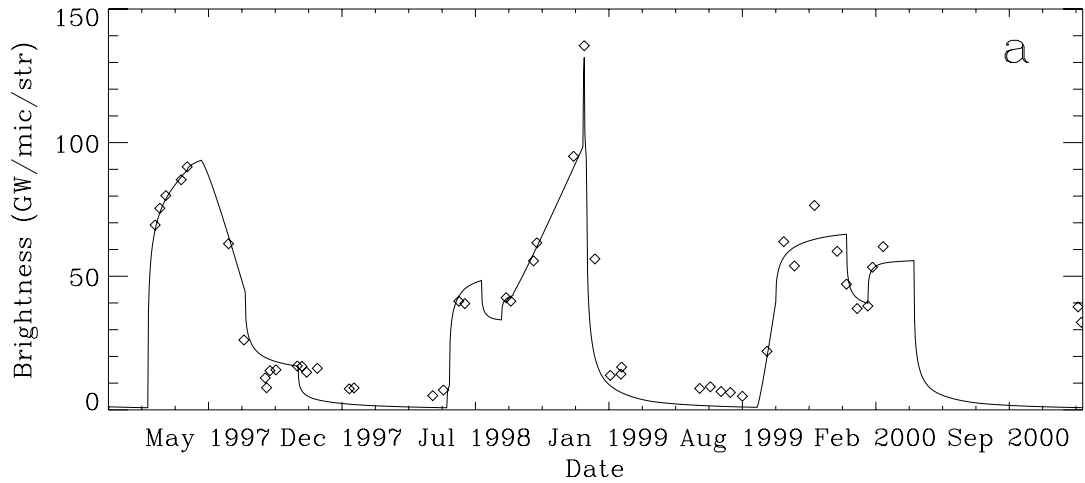
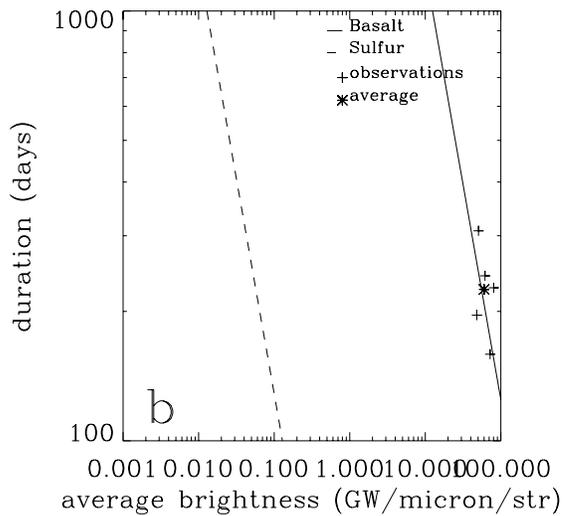
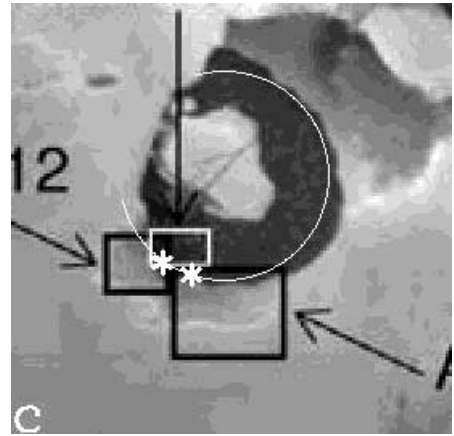

**Figure 3: Model results. a)** The solid line shows the modeled 3.5 μm brightness as a function of time for the period from May 1997 through September 2000. The diamonds are observed brightnesses. **b)** Duration and average brightnesses possible from model for both basalt and sulfur. Plus signs show the duration and average brightness for each of the five observed brightening events. The asterisk shows the average duration and brightness for all observed events. **c)** The locations of overturn front on July 12[th] and August 4[th], 1998 (asterisks) based on our model. The black boxes show the position of the hottest material seen on those dates by MacIntosh et al. (2003). The horseshoe shows the long dimension of the model wrapped around the patera. Bacground image from MacIntosh et al. (2003).

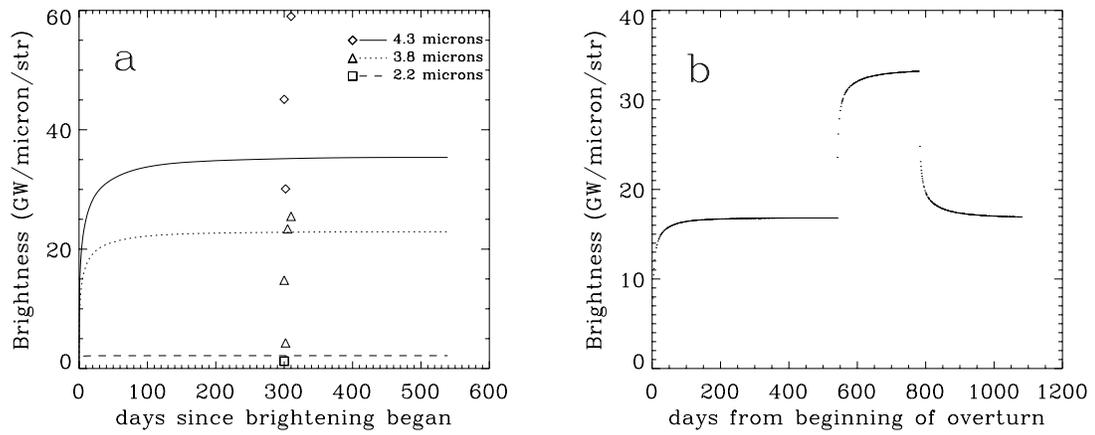

**Figure 4: Results of model for 2001 through 2004. a)** Brightness of Loki in 2001 as measured by Marchis et al. (2005) at different wavelengths. The lines are modeled brightnesses at those wavelengths assuming a speed of 0.5 km/day. **b)** Modeled brightness as a function of time assuming a wave speed of 0.5 km/day and a second front beginning 540 days after the first.

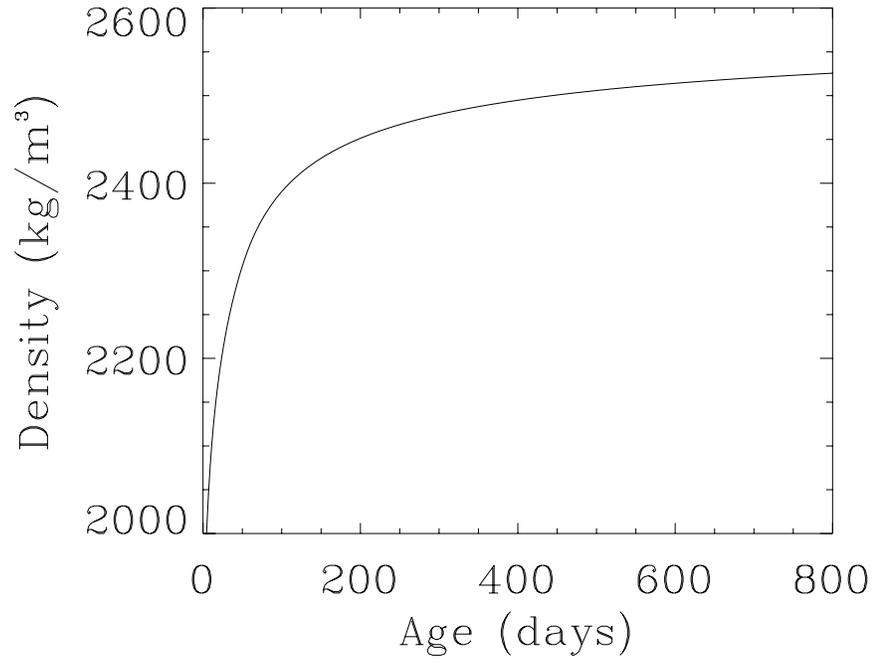

**Figure 5:** The density of the solidified crust as a function of time assuming the porosity profile of Peck et al. (1966).